# Fast computation of kernel statistics using genotype value decomposition

Kazuharu Misawa*

*Corresponding author

Email: kazu_misawa@hotmail.com

Kansai Medical University,

2-5-1 Shin-machi, Hirakata, Osaka, Japan 573-1010

Phone: +81-72-804-2623

Fax: +81-72-804-2630

1,356 words

# Abstract

**Because of the recent advances of genome sequences, a large number of human genome sequences are available for the study of human genetics. Genome-wide association studies typically focus on associations between single-nucleotide polymorphisms and traits such as major human diseases. However, the statistical power of classical single-marker association analysis for rare variants is limited. To address the challenge, rare and low-frequency variants are often grouped into a gene or pathway level, and the effects of multiple variants evaluated based on collapsing methods. The sequential kernel association test (SKAT) is one of the most effective collapsing methods. SKAT utilizes the kernel matrix. The size of the kernel matrix is $O(n^2)$, where the sample size is n, so that the calculation of the data using the kernel method requires a long time. As the sample sizes of human genetic studies increase, the computational time is getting more and more problematic. In the present paper, the genotype value decomposition method is proposed for the handling the sequential kernel in a short period of time. The method can be referred to as genotype value decomposition. In the present paper, it is shown that the genetic relationship matrix and Identity by State (IBS) matrix can be obtained using the genotype value vectors. By using this method, the SKAT can be conducted with time complexity $O(n)$. The proposed method enables to conduct SKAT for samples of human genetics.**

# Introduction

Genome-wide association studies typically focus on associations between single-nucleotide polymorphisms and traits such as major human diseases. However, the statistical power of classical single-marker association analysis for rare variants is limited. To address the challenge, rare and low-frequency variants are often grouped into a gene or pathway level, and the effects of multiple variants evaluated based on collapsing methods. The sequential kernel association test (SKAT) (WU *et al.* 2011) is one of the most effective collapsing methods (LARSON *et al.* 2019).

SKAT applies a test statistic $Q$, which is defined by the quadratic form, $Q = \boldsymbol{y}^{\mathrm{T}}\mathbf{K}\boldsymbol{y}$, where $\boldsymbol{y}$ is the zero-centered column vector of the phenotype. Namely, the average of the elements of $\boldsymbol{y}$ is 0. Since the length of $\boldsymbol{y}$ is $n$ and the size of the matrix $\mathbf{K}$ is $n^2$, the computational time for calculating $Q$ is $O(n^2)$. The size of the kernel matrix is $\mathrm{O}(n^2)$ where $n$ is the sample size. When the number of people is 10,000, the size of the kernel matrix is 100,000,000.

The aim of the present study is to obtain the algorithm for obtaining $Q$ in a short time. I refer to the method as "genotype value decomposition." By using the genotype value decomposition, calculation of the statistics $Q$ can be reduced to time complexity $O(n)$.

# Algorithm for genotype value decomposition

## *Genotype value vectors*

In the present study, all sites are assumed to be biallelic, namely, each site has a reference allele and an alternative allele. The allele values are 0 for the reference allele, 1 for the alternative allele. The separator between the alleles is "/" as used in the variant call format (DANECEK *et al.* 2011).Let $g_k(i) \in \{0,1,2\}$ be the genotype value for individual $i$ at locus $k$. In the present study, $g_k(i)$ is the number of alternative alleles. The relationship between the genotype of the individual $i$ at locus $k$ and $g_k(i)$ is shown in Table 1. The vector $\boldsymbol{g}_k$ is defined by

$$\boldsymbol{g}_k = (g_k(1) \quad g_k(2) \quad \cdots \quad g_k(m))^{\mathrm{T}} \tag{1}$$

Let us define genotype value vectors at locus $k$. Let $a_k(i)$ be 1 when the individual $i$ is a homozygote of a reference allele at locus $k$, otherwize $a_k(i) = 0$. Let $b_k(i)$ be 1 when the individual $i$ is the heterozygote at locus $k$, otherwise $b_k(i) = 0$. Let $c_k(i)$ be 1 when the individual $i$ is a homozygote of an alternative allele at locus $k$, otherwize $c_k(i) = 0$. It is worth noting that $a_k(i) + b_k(i) + c_k(i) = 1$. The vectors $\boldsymbol{a}_k$, $\boldsymbol{b}_k$, $\boldsymbol{c}_k$, and $\boldsymbol{g}_k$ are also defined by

$$\boldsymbol{a}_k = (a_k(1) \quad a_k(2) \quad \cdots \quad a_k(m))^{\mathrm{T}}$$

$$\boldsymbol{b}_k = (b_k(1) \quad b_k(2) \quad \cdots \quad b_k(m))^{\mathrm{T}}$$

$$\boldsymbol{c}_k = (c_k(1) \quad c_k(2) \quad \cdots \quad c_k(m))^{\mathrm{T}} \tag{2}$$

Let us denote $\boldsymbol{a}_k$, $\boldsymbol{b}_k$, and $\boldsymbol{c}_k$ as the genotype value vectors. The alternative allele frequency, $p$, is obtained by $\left(\boldsymbol{b}_k{}^{\mathrm{T}}\mathbf{1} + 2\boldsymbol{c}_k{}^{\mathrm{T}}\mathbf{1}\right)/(2m)$ with computational time of $O(n)$, where $\mathbf{1}$ is defined by $\mathbf{1} = (1 \quad 1 \quad \cdots \quad 1)^{\mathrm{T}}$.

Table 1 displays the relationships among allelic states $a_k(i)$, $b_k(i), c_k(i)$, and $g_k(i)$. Since $g_k(i) = b_k(i) + 2c_k(i)$, $\boldsymbol{g}_k$ is obtained by $\boldsymbol{g}_k = \boldsymbol{b}_k + 2\boldsymbol{c}_k$ with computational time of $O(n)$.

*The genetic relationship matrix*

The genetic relationship matrix (GRM) among individuals is used in the software GCTA (YANG *et al.* 2011) and in principal component analysis (PRICE *et al.* 2006). Let us denote the GRM at locus $k$ as $\mathbf{X}_k$. We subtract the mean $\mu_k = \{\sum_{i=1}^{n} g_k(i)\}/n$ to obtain a matrix with row sums equal to 0. The $ij$-th element of $\mathbf{X}_k$ at locus $k$ is obtained using the following equation:

$$\mathbf{X}_k(i,j) = \{g_k(i) - \mu_k\}\{g_k(j) - \mu_k\}$$

$$= g_k(i)g_k(j) - \mu_k g_k(i) - \mu_k g_k(j) + {\mu_k}^2 \qquad (3)$$

Subsequently, the matrix $\mathbf{X}_k$ can be obtained using the genotype value vectors.

$$\mathbf{X}_k = \begin{pmatrix} g_k(1)g_k(1) & g_k(1)g_k(2) & \cdots & g_k(1)g_k(n) \\ g_k(2)g_k(1) & g_k(2)g_k(2) & \cdots & g_k(2)g_k(n) \\ \vdots & \vdots & & \vdots \\ g_k(n)g_k(1) & g_k(n)g_k(2) & \cdots & g_k(n)g_k(n) \end{pmatrix}$$

$$-\mu_k \begin{pmatrix} g_k(1) & g_k(1) & \cdots & g_k(1) \\ g_k(2) & g_k(2) & \cdots & g_k(2) \\ \vdots & \vdots & & \vdots \\ g_k(n) & g_k(n) & \cdots & g_k(n) \end{pmatrix} - \mu_k \begin{pmatrix} g_k(1) & g_k(2) & \cdots & g_k(n) \\ g_k(1) & g_k(2) & \cdots & g_k(n) \\ \vdots & \vdots & & \vdots \\ g_k(1) & g_k(2) & \cdots & g_k(n) \end{pmatrix}$$

$$+{\mu_k}^2 \begin{pmatrix} 1 & 1 & \cdots & 1 \\ 1 & 1 & \cdots & 1 \\ \vdots & \vdots & & \vdots \\ 1 & 1 & \cdots & 1 \end{pmatrix} \qquad (4)$$

The entire GRM is obtained using $\mathbf{X} = \sum_{k=1}^{l} w_k \mathbf{X}_k$, where $w_k$ is weight of locus $k$ and $l$ is the number of loci. Weights are usually put to normalize each data column to have the same variance (PATTERSON *et al.* 2006). To boost analysis power, SKAT allows the incorporation of flexible weight functions (WU *et al.* 2011). Putting weights does not affect the complexity of calculation,

Let us define a new matrix $\mathbf{G}_k$. As shown in table 1, the $ij$-th element of $\mathbf{G}_k$ at locus $k$ is can be calculated using the following equation:

$$\mathbf{G}_k(i,j) = g_k(i)g_k(j) \tag{5}$$

Subsequently, the matrix $\mathbf{G}_k$ would be obtained using the genotype value vectors.

$$\mathbf{G}_k = \begin{pmatrix} g_k(1)g_k(1) & g_k(1)g_k(2) & \cdots & g_k(1)g_k(m) \\ g_k(2)g_k(1) & g_k(2)g_k(2) & \cdots & g_k(2)g_k(m) \\ \vdots & \vdots & & \vdots \\ g_k(m)g_k(1) & g_k(m)g_k(2) & \cdots & g_k(m)g_k(m) \end{pmatrix}$$

$$= \begin{pmatrix} g_k(1) \\ g_k(2) \\ \vdots \\ g_k(m) \end{pmatrix} \begin{pmatrix} g_k(1) & g_k(2) & \cdots & g_k(m) \end{pmatrix} = \boldsymbol{g}_k \boldsymbol{g}_k^{\mathrm{T}} \tag{6}$$

Therefore,

$$\mathbf{X}_k = \mathbf{G}_k - \mu \mathbf{1} \boldsymbol{g}_k^{\mathrm{T}} - \mu \boldsymbol{g}_k \mathbf{1}^{\mathrm{T}} + \mu^2 \mathbf{1}\mathbf{1}^{\mathrm{T}} \tag{7}$$

It is worth noting that

$$\boldsymbol{y}^{\mathrm{T}} \mathbf{X}_k \boldsymbol{y} = \boldsymbol{y}^{\mathrm{T}} \mathbf{G}_k \boldsymbol{y} + \mu \boldsymbol{y}^{\mathrm{T}} \mathbf{1} \boldsymbol{g}_k^{\mathrm{T}} \boldsymbol{y} + \mu \boldsymbol{y}^{\mathrm{T}} \boldsymbol{g}_k \mathbf{1}^{\mathrm{T}} \boldsymbol{y} + \mu^2 \boldsymbol{y}^{\mathrm{T}} \mathbf{1}\mathbf{1}^{\mathrm{T}} \boldsymbol{y} \tag{8}$$

Since $\boldsymbol{y}^{\mathrm{T}} \mathbf{1} = \mathbf{1}^{\mathrm{T}} \boldsymbol{y} = 0$, we obtain

$$\boldsymbol{y}^{\mathrm{T}} \mathbf{X}_k \boldsymbol{y} = \boldsymbol{y}^{\mathrm{T}} \mathbf{G}_k \boldsymbol{y}. \tag{9}$$

By using distributivity and associativity of matrix production, we obtain

$$\mathbf{G}_k = (\boldsymbol{b} + 2\boldsymbol{c})(\boldsymbol{b} + 2\boldsymbol{c})^{\mathrm{T}}.$$

$$Q_k = \mathbf{y}^{\mathrm{T}}\mathbf{G}_k\mathbf{y} = \mathbf{y}^{\mathrm{T}}(\boldsymbol{b} + 2\boldsymbol{c})(\boldsymbol{b} + 2\boldsymbol{c})^{\mathrm{T}}\mathbf{y}$$

$$= (\mathbf{y}^{\mathrm{T}}\boldsymbol{b} + 2\mathbf{y}^{\mathrm{T}}\boldsymbol{c})(\mathbf{y}^{\mathrm{T}}\boldsymbol{b} + 2\mathbf{y}^{\mathrm{T}}\boldsymbol{c})^{\mathrm{T}}, \tag{10}$$

because the transpose of a product of matrices is the product, in the reverse order, of the transposes of the factors. Note that $\mathbf{y}^{\mathrm{T}}\boldsymbol{b}$ and $\mathbf{y}^{\mathrm{T}}\boldsymbol{c}$ are scholars so that they can be obtained using the inner product of two vectors with a computational time of $O(n)$. For SKAT, Wu et al. (WU *et al.* 2011) defined a test statistic $Q$ as $Q = \boldsymbol{y}^{\mathrm{T}}\mathbf{X}\boldsymbol{y}$. $Q$ can be obtained using $Q = \sum_{k=1}^{l} w_k Q_k$.

*IBS kernel*

IBS defines similarity between individuals as the number of shared alleles. The IBS kernel is used in linear regression (WESSEL AND SCHORK 2006; KWEE *et al.* 2008) and the SKAT (WU *et al.* 2011). Wu et al. (WU *et al.* 2011) mentioned that the triangular kernel (FLEURET AND SAHBI 2003) is the same as the IBS kernel. Let $\mathrm{IBS}_k(i) \in \{0,1,2\}$ be the $ij$-th element of the IBS matrix, $\mathbf{IBS}_k$, at locus $k$, denotes the number of shared alleles by subjects $i$ and $j$ at locus $k$. Table 2 displays the relationships among genotype values and IBS. From the table, we can observe the following relationship among genotype value vectors and the IBS matrix.

$$\mathrm{IBS}_k(i,j) = 2a_k(i)a_k(j) + b_k(i) + b_k(j) + 2c_k(i)c_k(j). \tag{11}$$

Thus, IBS matrix at locus $k$ is obtained by

$$\mathbf{IBS}_k = 2\begin{pmatrix} a_k(1)a_k(1) & a_k(1)a_k(2) & \cdots & a_k(1)a_k(m) \\ a_k(2)a_k(1) & a_k(2)a_k(2) & \cdots & a_k(2)a_k(m) \\ \vdots & \vdots & & \vdots \\ a_k(m)a_k(1) & a_k(m)a_k(2) & \cdots & a_k(m)a_k(m) \end{pmatrix}$$

$$+\begin{pmatrix} b(1) & b(1) & \cdots & b(1) \\ b(2) & b(2) & \cdots & b(2) \\ \vdots & \vdots & & \vdots \\ b(m) & b(m) & \cdots & b(m) \end{pmatrix} + \begin{pmatrix} b(1) & b(2) & \cdots & b(m) \\ b(1) & b(2) & \cdots & b(m) \\ \vdots & \vdots & & \vdots \\ b(1) & b(2) & \cdots & b(m) \end{pmatrix}$$

$$+2\begin{pmatrix} c(1)c(1) & c(1)c(2) & \cdots & c(1)c(m) \\ c(2)c(1) & c(2)c(2) & \cdots & c(2)c(m) \\ \vdots & \vdots & & \vdots \\ c(m)c(1) & c(m)c(2) & \cdots & c(m)c(m) \end{pmatrix}$$

$$= 2\boldsymbol{a}_k\boldsymbol{a}_k{}^{\mathrm{T}} + \mathbf{1}\boldsymbol{b}_k{}^{\mathrm{T}} + \boldsymbol{b}_k\mathbf{1}^{\mathrm{T}} + 2\boldsymbol{c}_k\boldsymbol{c}_k{}^{\mathrm{T}}. \qquad (12)$$

By using distributivity and associativity of matrix production, we obtain

$$\boldsymbol{y}^{\mathrm{T}}\mathbf{IBS}_k\boldsymbol{y} = \boldsymbol{y}^{\mathrm{T}}\left(\boldsymbol{a}_k\boldsymbol{a}_k{}^{\mathrm{T}} + \mathbf{1}\boldsymbol{b}_k{}^{\mathrm{T}} + \boldsymbol{b}_k\mathbf{1}^{\mathrm{T}} + \boldsymbol{c}_k\boldsymbol{c}_k{}^{\mathrm{T}}\right)\boldsymbol{y}$$

$$= 2\boldsymbol{y}^{\mathrm{T}}(\boldsymbol{a}\boldsymbol{a}^{\mathrm{T}})\boldsymbol{y} + 2\boldsymbol{y}^{\mathrm{T}}(\boldsymbol{c}\boldsymbol{c}^{\mathrm{T}})\boldsymbol{y} + \boldsymbol{y}^{\mathrm{T}}(\mathbf{1}\boldsymbol{b}^{\mathrm{T}})\boldsymbol{y} + \boldsymbol{y}^{\mathrm{T}}(\boldsymbol{b}\mathbf{1}^{\mathrm{T}})\boldsymbol{y}$$

$$= 2(\boldsymbol{y}^{\mathrm{T}}\boldsymbol{a})(\boldsymbol{a}^{\mathrm{T}}\boldsymbol{y}) + 2(\boldsymbol{y}^{\mathrm{T}}\boldsymbol{c})(\boldsymbol{c}^{\mathrm{T}}\boldsymbol{y}). \qquad (13)$$

$\boldsymbol{y}^{\mathrm{T}}\boldsymbol{a}$ and $\boldsymbol{y}^{\mathrm{T}}\boldsymbol{c}$ are scholars and they can be obtained using the inner product of two vectors with computational time of $O(n)$.

## Results and Discussion

In the present paper, the author demonstrates that the genetic relationship matrix and Identity by State (IBS) matrix are obtained using the genotype value vectors. Consequently, the SKAT can be conducted with time complexity $O(n)$ where $n$ is the sample size. The method can be referred to as genotype value decomposition.

Hasegawa et al (HASEGAWA *et al.* 2016) obtained the p-value of SKAT by applying permutation procedures for SKAT. However, resampling requires a huge amount of computation time to obtain accurate p-values. Therefore, the proposed procedure is applicable to large amounts of data.

## Acknowledgments

This work was supported by Kakenhi (JP17K08682 and JP19K22647).

## Author's contributions

KM wrote the manuscript. KM read and approved the final manuscript.

## Declaration of Interests

The authors declare no competing interests.

Table 1. Relationship between genotype values and the number of alternative alleles.

| Genotype | $a_k(i)$ | $b_k(i)$ | $c_k(i)$ | $g_k(i)$ |
|---|---|---|---|---|
| 0/0 | 1 | 0 | 0 | 0 |
| 0/1 | 0 | 1 | 0 | 1 |
| 1/1 | 0 | 0 | 1 | 2 |

Table 2. Relationship between genotype values and identity by states (IBS)

| Individual $i$ | | | | Individual $j$ | | | | |
|---|---|---|---|---|---|---|---|---|
| Genotype | $a_k(i)$ | $b_k(i)$ | $c_k(i)$ | Genotype | $a_k(j)$ | $b_k(j)$ | $c_k(j)$ | IBS |
| 0/0 | 1 | 0 | 0 | 0/0 | 1 | 0 | 0 | 2 |
| 0/0 | 1 | 0 | 0 | 0/1 | 0 | 1 | 0 | 1 |
| 0/0 | 1 | 0 | 0 | 1/1 | 0 | 0 | 1 | 0 |
| 0/1 | 0 | 1 | 0 | 0/0 | 1 | 0 | 0 | 1 |
| 0/1 | 0 | 1 | 0 | 0/1 | 0 | 1 | 0 | 2 |
| 0/1 | 0 | 1 | 0 | 1/1 | 0 | 0 | 1 | 1 |
| 1/1 | 0 | 0 | 1 | 0/0 | 1 | 0 | 0 | 0 |
| 1/1 | 0 | 0 | 1 | 0/1 | 0 | 1 | 0 | 1 |
| 1/1 | 0 | 0 | 1 | 1/1 | 0 | 0 | 1 | 2 |